\documentclass[fleqn,10pt]{wlscirep}
\title{Atomic coherence effects in few-cycle pulse induced ionization}

\author[1,*]{Viktor Ayadi}
\author[2]{Mih\'aly G. Benedict}
\author[1,3]{P\'eter Dombi}
\author[2,3]{P\'eter F\"oldi}
\affil[1]{MTA "Lend\"ulet" Ultrafast Nanooptics Group, Wigner Research Centre for Physics, Konkoly-Thege M. \'ut 29-33, H-1121 Budapest, Hungary}
\affil[2]{Department of Theoretical Physics, University of Szeged, Tisza Lajos k\"or\'ut 84, 	H-6720 Szeged, Hungary}
\affil[3]{ELI-ALPS, ELI-HU Non-profit Ltd., Dugonics t\'er 13, H-6720 Szeged, Hungary}
\affil[*]{ayadi.viktor@wigner.mta.hu}

\begin{abstract}
The interaction of a short, few-cycle light pulse and an atom which is prepared initially in a superposition of two stationary states is shown to exhibit strong signatures of atomic coherence. For a given waveform of the laser pulse, appropriate quantum mechanical relative phase between the constituents of the initial superposition can increase the ionization probability by a factor of three. A similarly strong effect can be observed when the waveform of the ionizing pulse is changed. These results allow for intuitive explanations, which are in agreement with the numerical integration of the time dependent Schr\"{o}dinger equation.
\end{abstract}
\begin{document}

\flushbottom
\maketitle
%
%
\thispagestyle{empty}


\section*{Introduction}

During the process of photoionization induced by few-cycle, near infrared
laser pulses, both the period of the external field oscillations and the
duration of the complete pulse are on the same timescale as the internal
atomic dynamics. This suggests that the details of the electron emission are
determined by the waveform of the exciting laser pulse, in other words,
besides the temporal envelope of the pulse, the carrier-envelope phase (CEP) \cite{KB98,RE07}
also plays an important role in the process. Similar CEP-dependent effects
were shown to appear in various laser-induced processes
including high harmonic generation
\cite{BohanPhysRevLett98,Baltuska2003,SansonePRL2003}, above-threshold
ionization (ATI) \cite{PaulusRPL03,Paulus2001}, non-sequential double
ionization \cite{RottkePRL04} and multi-photon induced photoemission
\cite{Apolonski2004,DombiNewJPhys04}.

Recently, the excitation and ionization probabilities from ground state atomic
hydrogen have been studied in intense short pulses \cite{LiPRA14} by
numerically integrating the time dependent Schr\"{o}dinger equation (TDSE). The
interesting features found in that paper are related to the ratio of the ionization and excitation probabilities as a function of the intensity of the incoming laser field.
As it was found, whenever an ATI peak merges into the continuum as a consequence of the increasing ponderomotive force, the ionization yield has a minimum, while the excitation probability increases.
Let us note that phase relations play an important role in ATI.
This was pointed out in \cite{Nakajima07}, where the influence of a chirp
in the excitation pulse was investigated, as well as in
\cite{Hu:13}, where the effect of a sudden phase jump during the
excitation was considered.

In the current paper we also consider short pulse induced ionization
processes, but -- in contrast to previous studies -- we focus on the interplay between the internal atomic dynamics
and that induced by the intense laser field. To this end we consider
initial states that are superpositions of dipole coupled stationary bound
states. There is a great variety of important effects in atom-field
interactions, where atomic coherence between atomic levels play a role.
Coherent population trapping, electromagnetically induced transparency and
lasing without inversion, are the most significant ones in the normal
intensity regime. For an overview of all these effects see \cite{SZ97}.
Coherent superpositions as initial states in the context of high harmonic generation (HHG) have been
investigated in Refs. \cite{Averbukh, Gauthey, Watson}.
Atomic coherence plays also a decisive role in strong field atomic stabilization,
predicted first in \cite{FedorovMov88,PontGavrila90}, see also \cite{Fedorov97,Popov03}.
In order to observe this effect, however, pulses of duration of hundreds of
femtoseconds are necessary.

It is natural to expect that atomic coherence may influence the transitions to a
third energy level also in the field of an intense, few-cycle pulse, when the third level
falls in the continuum. We choose the initial conditions as coherent
superpositions of stationary atomic states in order to explore the details of this process.
The atom and the field are coupled
through the dipole interaction, and the expectation value of the dipole
moment operator in these initial states
is generally non-zero, it oscillates with the appropriate Bohr frequency. When
the frequency of these ("internal, atomic") oscillations is comparable with
that of the laser field, the essence of the pronounced CEP-dependence of the
problem can be understood using a simple, classical picture: The relative
phase of the internal oscillations and that of the external excitation
determines whether they interfere constructively or destructively. When the
"swing" of the internal dipole oscillations is excited in an appropriate
phase, the amplitude increases and the ionization yield has a maximum. On the
other hand, when the external field has to "work against" the internal
oscillations, the ionization probability is considerably lower. Additionally,
when the frequency of the two types of oscillations are not close enough, this
effect becomes weaker. Here, we will show that our fully quantum mechanical
treatment is qualitatively in accordance with the intuitive picture described above.

Clearly, it is not only the atomic coherence that determines the ionization probability: as one
can expect, the energy of the initial states is the most important parameter. As we shall see,
the ionization yield is not increasing monotonically with the expectation value of the energy in the
initial state, which can be understood by inspecting the distance of the first ATI peak from the limit of the continuum.
These effects define the ionization probability on which oscillations caused by atomic coherence
phenomena are superimposed.

Technically, we solve the time dependent Schr\"{o}dinger equation of a
hydrogen-like atom in an intense, pulsed, few-cycle, near infrared laser field.
Our approach is based on the numerical methods that were used e.g., in
Refs.~\cite{Krause92, MullerEfficient}. We use realistic parameters for the
high-power ultrashort laser pulses with durations as short as a few optical
cycles that have already been available as research tools
\cite{Nisoli96, Cavelieri, DombiOptExp05}. The results show strong CEP-dependence
of the ionization signal for initial superpositions of \textit{s} and \textit{p} states with different principal quantum numbers.
Additionally, we study how the relative phase between the components of
these superpositions influences the dynamics of the atomic dipole moment and
the ionization process. We show that the CEP dependence of the ionization
process can be detectable for pulse durations up to $22 \,\mathrm{fs}$ (8 optical cycles),
which is remarkably long in this context. Note that if
the preparation of the initial superpositions is assumed to be achieved by a
conventional, many cycle pulse of appropriate area \cite{AE75}, the relative
phase of the \textit{s} and \textit{p} components is proportional to the delay between the
preparation and the few-cycle pulses. As we discuss in the second and the fourth sections, the time scales of spontaneous emission,  as well as that of the “conventional,” low intensity, many cycle excitation compared with the ultrashort ionization allows a convenient way of preparing the initial superpositions.

The current paper is organized as follows. In the second section we introduce
the specific initial states we consider, describe the excitation pulse with a
carrier-envelope phase, as well as the method of integrating the TDSE. Next we present and discuss
our results, and analyse the possible experimental realization. The conclusions can be found in the final section.


\section*{Model}
\label{modelsec}

In order to focus on the interplay between the internal, atomic dynamics and
that of induced by the strong laser pulse, we consider three types of initial
superpositions of hydrogenic states
\begin{align}
\psi_{{23}}(c,\delta)  &  =c_{2s}\phi_{{2s}}+c_{3p}\exp
(i\delta)\phi_{{3p}},\label{psiI}\\
\psi_{{34}}(c,\delta)  &  =c_{3s}\phi_{{3s}}+c_{4p}\exp
(i\delta)\phi_{{4p}},\label{psiII}\\
\psi_{{45}}(c,\delta)  &  =c_{4s}\phi_{{4s}}+c_{5p}\exp
(i\delta)\phi_{{5p}}, \label{psiIII}%
\end{align}
where $\phi_{n\ell}$ are normalized nonrelativistic hydrogenic eigenstates
with magnetic quantum number $m=0$. The coefficients $c_{n\ell}$ denote the real amplitudes of the
corresponding components in the superpositions, while $\delta$ is the relative
phase between them. Note that for the realistic laser parameters
we use [see after Eq.~\ref{Efield}], superpositions
of the $1s$ and $2p$ states lead to negligible ionization probability.

The time dependence of the nonstationary states (\ref{psiI} -
\ref{psiIII}) are determined
by the difference between the energy eigenvalues, i.e., the Bohr frequencies
$\omega_{nn^{\prime}}=\omega_{n}-\omega_{n^{\prime}}=
20,671\left(  \frac{1}{n^{\prime2}}-\frac{1}{{n}^{2}}\right)  (\mathrm{fs})^{-1}$ as:
\begin{equation}
\psi_{{nn^{\prime}}}(c,\delta,t)=e^{-i\omega_{n}t}\left\{  c_{ns}%
\phi_{{ns}}+c_{n^{\prime}p}\exp[i(\delta+\omega_{nn^{\prime}}%
t)]\phi_{{n^{\prime}p}})\right\}  , \label{evol0}%
\end{equation}
where the global phase factor has no physical consequences. We assume that
this is the wavefunction of the state until the strong external driving is
switched on.  The initial superpositions $\psi_{{nn^{\prime}}}(c,\delta,t=0)$
can be realized by using resonant, many cycle preparation pulses of low intensity, where the pulse area
(that is proportional to the time integral of the electric field amplitude)
determines the magnitude of the coefficients $c$ \cite{AE75}.
E.g., the state $\psi_{{23}}$ can be achieved  by an ultraviolet excitation of the 1s-3p
Lyman beta transition followed by a laser pulse of appropriate area at the  wavelength of the
Balmer alpha line in the visible. Another possibility is to start
the coherent superposition from the lower, metastable  2s state, which can be achieved
by various techniques, as described in \cite{Yatsenko}.
Note that the duration of the preparation process is still considerably shorter than the lifetime of the states (\ref{psiI} -
\ref{psiIII}) \cite{Jitrik}. For more details see the fourth section.
As it is suggested by the second term in Eq.~(\ref{evol0}),
the relative phase $\delta$ in Eqs.~(\ref{psiI} -\ref{psiIII}) can be controlled by changing the delay between the preparation pulse and
the few-cycle one.

Our main purpose is to investigate how short, intense, few-cycle laser pulses
modify the free time evolution given by Eq.~(\ref{evol0}) -- in a parameter
range where the duration of the external disturbance is comparable with the
time scale of the Bohr oscillations. The electric field of the few-cycle laser
pulse is assumed to be polarized in the $z$ direction, and its time dependence
is written as
\begin{equation}
E(t)=E_{0}\sin^{2}\left(  {{\pi t}/{\tau}}\right)  \cos(\omega t+\varphi
_{\mathrm{CEP}}), \label{Efield}%
\end{equation}
where the envelope function ($\sin^{2}$) is assumed to be zero when $t<0$ or
$t>\tau$ \cite{BandraukPhysRevA04}, and $\omega$ is the angular frequency
corresponding to a central wavelength of $800 \ \mathrm{nm}.$
Additionally, unless stated otherwise, we use $\tau=12\, \mathrm{fs}$ [corresponding
to $4.4\,\mathrm{fs}$ intensity full width at half maximum (FWHM)] and $E_{0}=2.5\, \mathrm{GV/m}$ peak field strength. These parameters are experimentally achievable using
current Ti:sapphire technology \cite{Cavelieri,DombiOptExp05,KrauszRevModPhys09}, and they induce comparable
ionization probabilities from the initial states (\ref{psiI} -
\ref{psiIII}).

We use atomic units and solve the time dependent Schr\"{o}dinger equation
(TDSE) numerically in the coordinate representation:
\begin{equation}
i\frac{\partial}{\partial t}\psi(\mathbf{r},t)=H(t)\ \psi(\mathbf{r},t),
\label{TDSE}%
\end{equation}
where the Hamiltonian takes the form
\begin{equation}
H(\mathbf{r},t)=H_{0}(\mathbf{r})+H_{\mathrm{I}}(\mathbf{r},t)=\left[  -\frac{\Delta
}{2}-\frac{1}{r}\right] + E(t)z. \label{Ham}%
\end{equation}
{Note that the position vector $\mathbf{r}=(x,y,z)$ corresponds to the
relative (electron-nucleus) coordinates, i.e., we are working in the center of
mass frame. The light-atom interaction described by $H_{\mathrm{I}}$ }${=}E(t)z$ {is
written using the dipole approximation (which can be shown to be valid in the
parameter range considered here \cite{BandraukJPhysB13, ReissJPhysB14}).
}

Note that various numerical methods can be used for the solution of the
TDSE as a partial differential equation,
for example: the method of lines, split step Fourier, etc. For our purposes, the
most efficient approach was found to be based on spherical harmonics expansion
\cite{Krause92}. That is, we use spherical coordinates and write $\psi$ as
\begin{equation}
\psi(r,\theta,t)=\sum_{\ell=0}^{\ell_{\mathrm{max}}}\frac{\Phi_{\ell}(r,t)}%
{r}Y_{\ell}^{0}(\theta), \label{psiBase}%
\end{equation}
where, due to the cylindrical symmetry, no summation with respect to $m$ appears:
we can restrict our calculations to $m=0.$ (In other words, there is no
$\varphi$ dependence.) Note that Eq.~(\ref{psiBase}) means a separation of the
$1/r$ dependence of the wave function, resulting in the following equations
for $\Phi_{\ell}(r,t)$  and $\hat{\psi}=r\psi:$
\begin{equation}
\hat{H}_{0}{\Phi_{\ell}(r,t)}=\left[  -\frac{1}{2}\left(  \frac{\partial^{2}%
}{\partial r^{2}}-\frac{\ell(\ell+1)}{r^{2}}\right)  -\frac{1}{r}\right]  {\Phi
_{\ell}(r,t)}, \label{H_0}%
\end{equation}%
\begin{equation}
i\frac{\partial}{\partial t}\hat{\psi}(\mathbf{r},t)=\left[  \hat{H}_{0}%
+\hat{H}_{\mathrm{I}}(t)\right]  \ \hat{\psi}(\mathbf{r},t). \label{TDSE2}%
\end{equation}
Since $H_{\mathrm{I}}$ does not contain derivatives with
respect to $\mathbf{r},$ the interaction $\hat{H}_{\mathrm{I}}=H_{\mathrm{I}}$ contains only the
$z$ coordinate, the nonzero matrix elements of which are well known, and read
\begin{equation}
\langle Y_{\ell}^{0}|\cos\theta|Y_{\ell+1}^{0}\rangle={\frac{\ell+1}{\sqrt
{(2\ell+1)(2\ell+3)}}}. \label{c_l}%
\end{equation}
The radial equation above is discretized by a special finite difference (FD)
scheme, which was presented in details in Ref.~\cite{MullerEfficient} and
relies on the alternating direction implicit (ADI) method \cite{PeacemanADI}.

For our calculations $\ell_{\mathrm{max}}$ was chosen to be $100$ and the radial
grid consisted of $6000$ points. The size of the computational grid we have used
was 600 atomic units ($\approx 31.7$ nm) in both directions.
These numbers were found to be sufficient:
The populations of the states close to the maximal $\ell$ and $r$ values were
always negligible in our calculations, i.e., there were no numerical artifacts
do to "reflections" at the edges of the grid.
\begin{figure}[ptb]
\centering
\includegraphics[width=8cm]{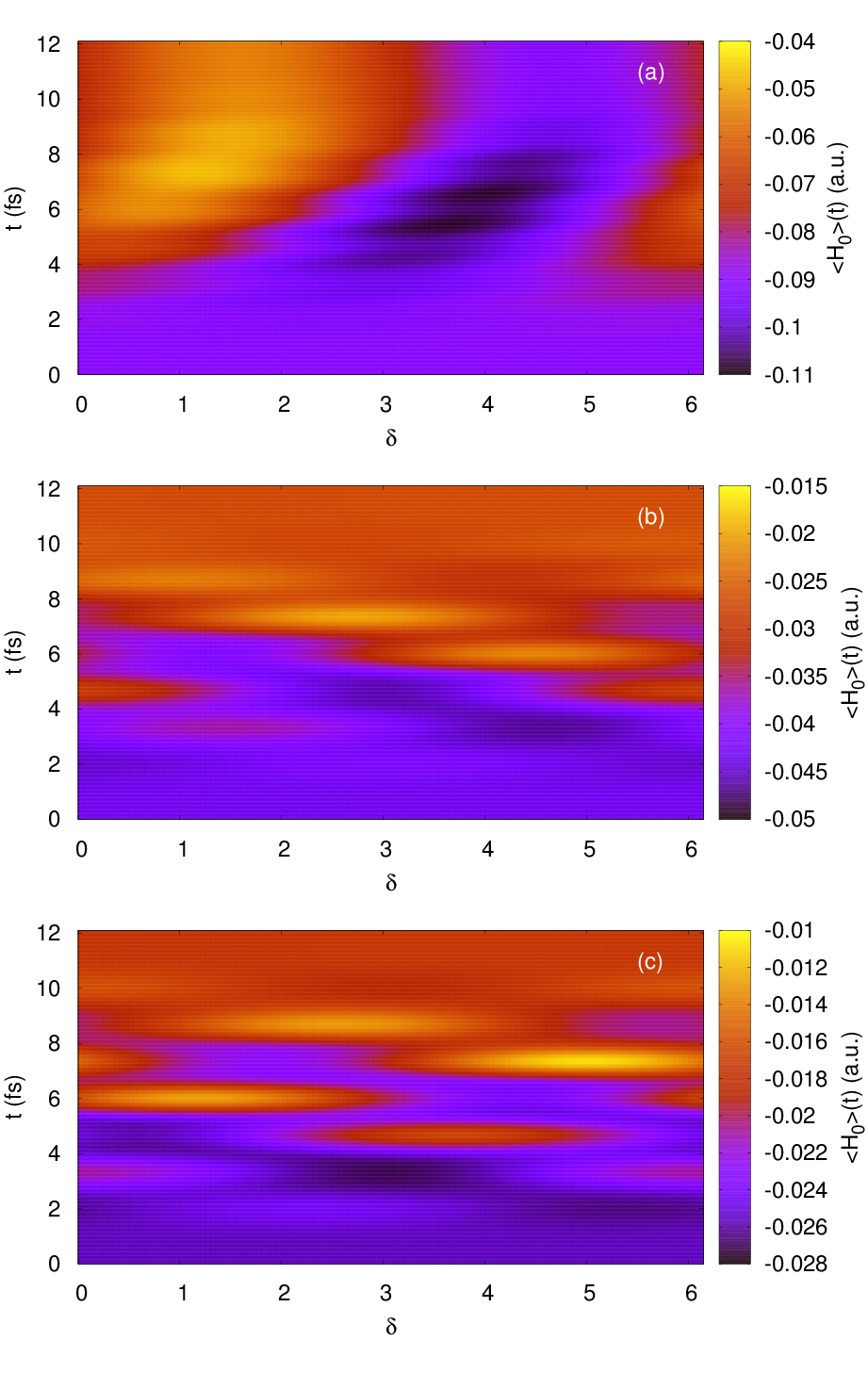}
\caption{The time and the initial phase
($\delta$) dependence of $\langle{H}_{0}\rangle(t)$ for the initial states: (a)
$\psi_{{23}}$, (b) $\psi_{{34}}$, (c) $\psi_{{45}},$ with $c_{ns}=c_{n+1 p}=1/\sqrt{2}$ for all cases.
The CEP of the laser pulse was chosen to be $0$ for all three states. Additional
parameters: $800\,\mathrm{nm}$ central wavelength, $\tau=12\,\mathrm{fs}$ and
$E_{0}=2.5\,\mathrm{GV/m}.$}%
\label{fig1}%
\end{figure}

\begin{figure}[ptb]
\centering
\includegraphics[width=8cm]{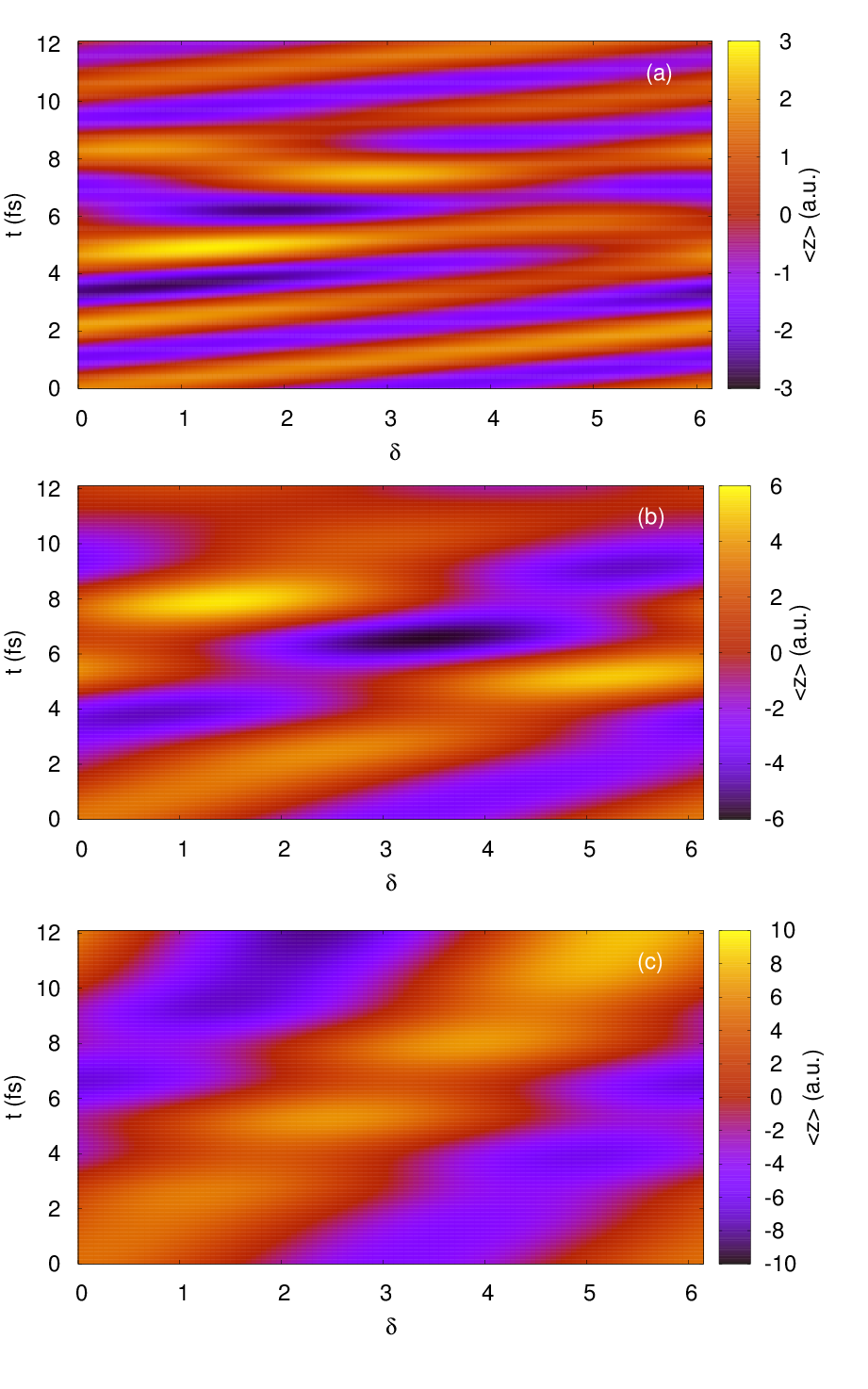}
\caption{The time and the initial phase
($\delta$) dependence of $\langle z\rangle$ for the initial states: (a)
$\psi_{{23}}$, (b) $\psi_{{34}}$, (c) $\psi_{{45}}$ (equal weight superpositions, $c_{ns}=c_{n+1 p}=1/\sqrt{2}$). The
parameters of the exciting pulse are the same as in Fig.~\ref{fig1} }%
\label{fig2}%
\end{figure}

The discretization of Eq.~(\ref{TDSE2}) allows the numerical determination
of the eigenvectors $|\Psi_{n}\rangle$ and eigenvalues $\epsilon_{n}$ of
${H}_{0}.$ The analytically known bound part of the spectrum [states with
negative energies, like $\phi_{{ns}}$ and $\phi_{{n
p}}$ in Eqs.~(\ref{psiI} -\ref{psiIII})] were appropriately reproduced in this way. States with
positive values of $\epsilon_{n}$ correspond to the continuous part of the
spectrum. Using a projector
\begin{equation}
\Pi_{\mathrm{i}}=\sum_{n,\epsilon_{n}>0}|\Psi_{n}\rangle\langle\Psi_{n}|
\end{equation}
on these states, the ionization probability can be defined as:
\begin{equation}
P_{\mathrm{i}}(t)=||\Pi_{\mathrm{i}}\psi(\mathbf{r},t)||^{2}=\sum_{n,\epsilon_{n}>0}|\langle
\Psi_{n}|\psi(\mathbf{r},t)\rangle|^{2}. \label{dissprob}%
\end{equation}
Note that since the complete state is normalized in the sense above ($\sum
_{n}|\langle\Psi_{n}|\psi(\mathbf{r},t)\rangle|^{2}=1$, within numerical
precision), $P_{\mathrm{i}}$ defined in Eq.~(\ref{dissprob}) has indeed a clear
probabilistic interpretation,
and after the pulse (i.e., for $t>\tau$), $P_{\mathrm{i}}$ can be determined experimentally.

\section*{Results and discussion}
\label{resultsec}
\subsection*{The role of the atomic coherence}
First we concentrate on superpositions with equal weights of the constituents [i.e., all the coefficients $c$ are equal to $1/\sqrt{2}$ in Eqs.~(\ref{psiI} -
\ref{psiIII})] and investigate the dependence of the ionization process on the relative phase $\delta.$
During the solution of the TDSE, we calculated the expectation value $\langle
z \rangle,$ which is proportional to the $z$ component of the dipole moment
(all other components of which are constant zero as a consequence of the
cylindrical symmetry). Although our system is not closed, and consequently
energy is not conserved, $\langle{H}_{0}\rangle(t)$ provides information
related e.g., to the population of the energy levels. The ionization
probability $P_{\mathrm{i}}$ measures the population of energy levels with positive energy.

\begin{figure}[ptb]
\centering
\includegraphics[width=8cm]{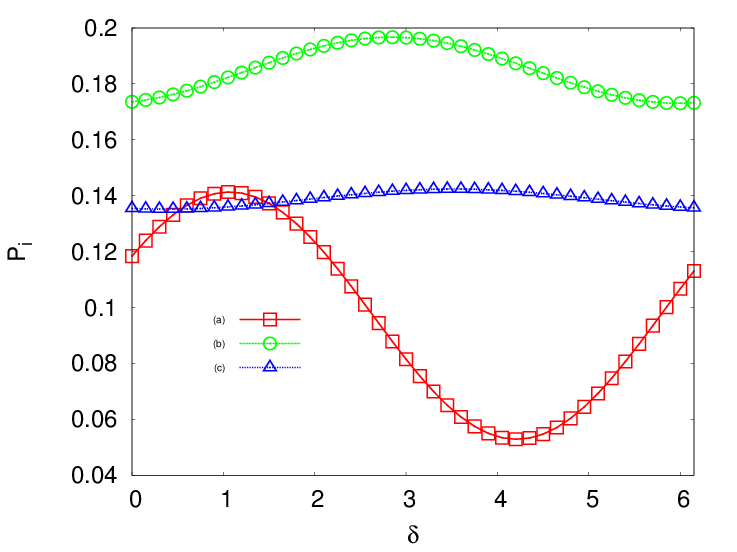}
\caption{The phase ($\delta$) dependence
of the final ionization probabilities for the initial states: (a) $\psi_{{23}}$,
(b) $\psi_{{34}}$, (c) $\psi_{{45}}$ (equal weight superpositions). The CEP of the laser pulse
was chosen to be $0$ for all three states. The parameters of the exciting
pulse are the same as in Fig.~\ref{fig1}.}%
\label{fig3}%
\end{figure}

Now, let us focus on Figs.~\ref{fig1}-\ref{fig3}, where $\varphi
_{\mathrm{CEP}}=0$ is fixed and we only varied $\delta$. Fig.~\ref{fig1} shows
$\langle{H_{0}}\rangle,$ for the three different initial superpositions
(\ref{psiI}-\ref{psiIII}). The general feature -- independently also from the
value of $\delta$ -- is that $\langle{H_{0}}\rangle$ does nor increase
monotonically, there are oscillations mainly at the laser carrier frequency.
This corresponds to the periodic motion of the electronic wave packet closer
and further away from the nucleus, as it can be seen in Fig.~\ref{fig2}, where
$\langle z \rangle$ is plotted. Note that when there is no external field
($t<0$ or $t>\tau$), the Hamiltonian does not contain time dependent terms
($H(t)=H_{0}$), and $\langle{H_{0}}\rangle$ becomes constant, which is not the
case for $\langle z \rangle,$ which oscillates already before the arrival of
the laser pulse and, in general, still oscillates when the pulse induced
external disturbance is over.

The ionization probability $P_{\mathrm{i}}$ also becomes constant for $t>\tau.$ By
inspecting Fig.~\ref{fig3}, we can immediately see that the final ionization
probabilities ($P_{\mathrm{i}}(\infty)=P_{\mathrm{i}}(t)$ for $t>\tau$) can be well approximated
by oscillating functions of the form
\begin{equation}
I(\delta)=A\sin(\delta+\delta_{I})+B. \label{I_delta}%
\end{equation}
Fitting the actual data, we get $A_{{23}}=0.130, \, B_{{23}%
}=0.110$, $A_{{34}}=0.010, \, B_{{34}}=0.185$ and
$A_{{45}}=0.010, \, B_{{45}}=0.150$ for panels a), b) and c),
respectively. These parameters show that the final ionization probability
exhibits the strongest $\delta$ dependence in the case of Fig.~\ref{fig3} a).
In other words, the dynamics of the states (\ref{psiI}) are extremely
sensitive to the initial phase. This phenomenon can be understood by recalling
that for the $2s-3p$ initial states the carrier frequency $\omega$ of the
incoming laser field is close to the Bohr frequency of the transition between
the states composing the superposition (the transition wavelength is $656.3\,
\mathrm{nm}$). The wavelengths of other transitions are more detuned (for $3s$
and $4p$ it is $1875\,\mathrm{nm}$ and for $4s$ and $5p$ it is
$4051\,\mathrm{nm}$), resulting in weaker $\delta$ dependence.
(We shall return to the analysis of the average value around which the final ionization probability oscillates in the next subsection.)
The relation of the Bohr frequencies and the carrier
frequency of the external field can be seen also in Fig.~\ref{fig2}, where the
most pronounced oscillations are related to $\omega_{nn^{\prime}},$ while the
perturbations change with $\omega,$ i.e., the central frequency of the laser
pulse. As we can see, the period of the two types of oscillations are the
closest for the $2s-3p$ initial superposition.

\begin{figure}[ptb]
\centering
\includegraphics[width=8cm]{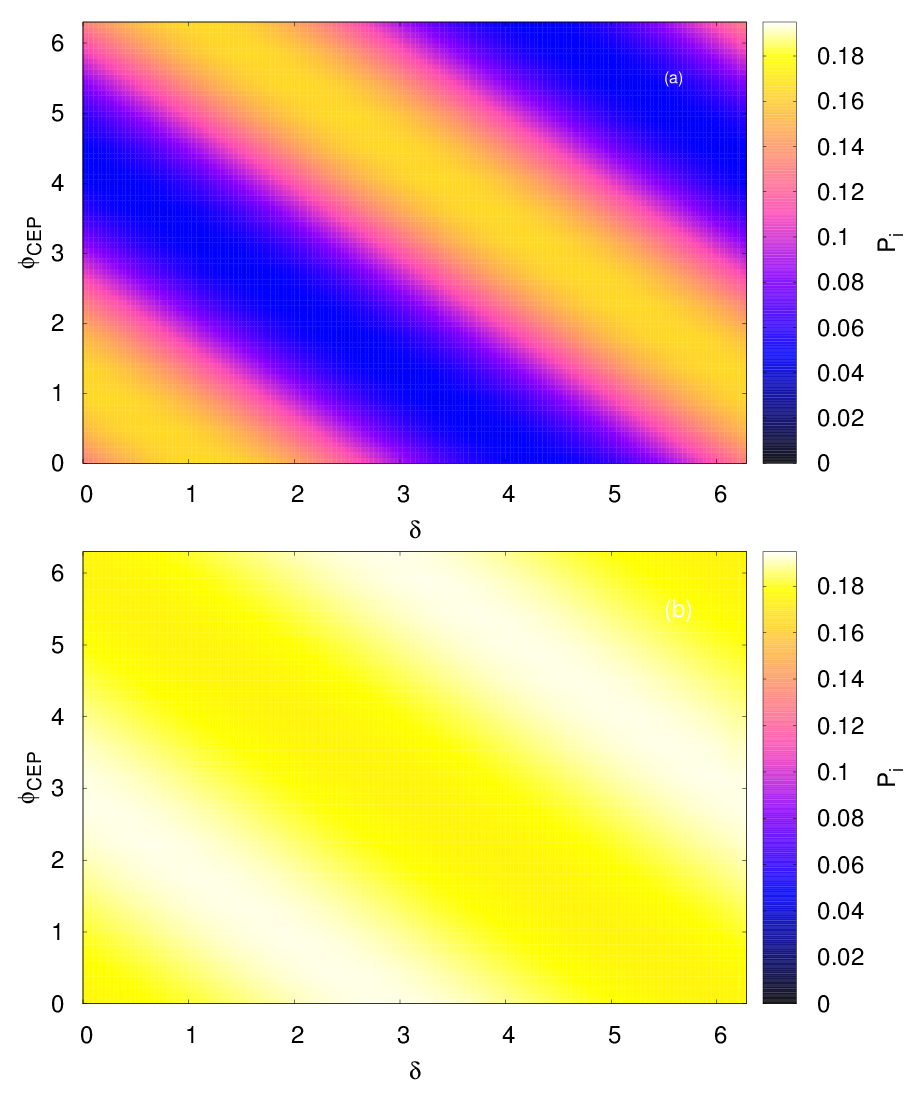}
\caption{The CEP and the initial phase
($\delta$) dependence of the final ionization probabilities for the initial states (a) $\psi_{{23}}$ and (b) $\psi_{{34}}$ (equal weight superpositions). The parameters of the
exciting pulse are the same as in Fig.~\ref{fig1}.}%
\label{fig4}%
\end{figure}

The effects shown in Figs.~\ref{fig1}-\ref{fig3} can be explained in a
simplified, but very intuitive manner: Without external disturbance, the $z$
component of the dipole moment would oscillate with the appropriate Bohr
frequency, $\omega_{nn^{\prime}}.$ A weak, almost resonant external field
pumps energy into the system and increases the amplitude of the oscillations.
However, when the external field is a short pulse containing a few optical
cycles only, the relative phase of the free dipole oscillations and that of
the excitation is important: the external field can both increase and decrease
the amplitude of the dipole oscillations. Clearly, for pulses that can induce
photoionization, strong nonlinear effects appear, numerous states can get
excited besides the initial one, etc., so the simple reasoning above can only
be relevant around the beginning of the pulse. However, this intuitive picture
as an initial approximation points out the role of the interplay between
internal, atomic dynamics and the excitation.

More details of this interplay can be investigated by changing the CEP of the
exciting pulse as well (which is possible also in an experimental
realization). The results are shown in Fig.~\ref{fig4},
where the final ionization probabilities are compared for the initial
states $\psi_{{23}}$ and
$\psi_{{34}}$ as a function of $\delta$ and $\varphi_{\mathrm{CEP}}$.
Here we can also see that for the nearly resonant case ($\psi_{{23}}$),
considerably more pronounced oscillatory behaviour occurs. More concretely,
$P_{\mathrm{i}}(t=\infty,\delta,\varphi_{\mathrm{CEP}})$ can again be fitted by
sinusoidal functions, in a way that their argument is $\delta+\varphi
_{\mathrm{CEP}}.$ The amplitude of the oscillations is more than ten times
larger for the initial states $\psi_{{23}}$ than for $\psi
_{{34}}.$

Since although few-cycle sources are available for experimental purposes, but
the shorter a pulse is, the more challenging its production is, it is worth
investigating whether the phase dependence of the problem is visible for
pulses longer than the one we considered so far. To this end, we performed
calculations for five longer pulses ($6.6$--$22$ $\mathrm{fs}$ FWHM in
the intensity) with the same peak intensity. The results are shown in
Fig.~\ref{fig6}. Obviously, longer pulses with the same peak intensity can pump more
energy into the atomic system, resulting in higher final ionization probabilities.
More interestingly, as we can see, the CEP dependence is still visible (and with
current experimental tools it should be also measurable) even for pulses as
long as $22\,\mathrm{fs}$ consisting of $8$ optical cycles, which is a
remarkable feature, since for typical strong-field interactions, CEP effects
disappear for pulses longer than $2-3$ optical cycles. Carrier-envelope phase
effects for multi-cycle pulses have only been observed by involving complex
extreme ultraviolet spectrscopic techniques in high harmonic generation \cite{Sansone04, Tzallas10}.
Additionally, besides the fundamental perspectives, this feature increases the number of laboratories where
our predictions can be tested experimentally.

\begin{figure}[ptb]
\centering
\includegraphics[width=8cm]{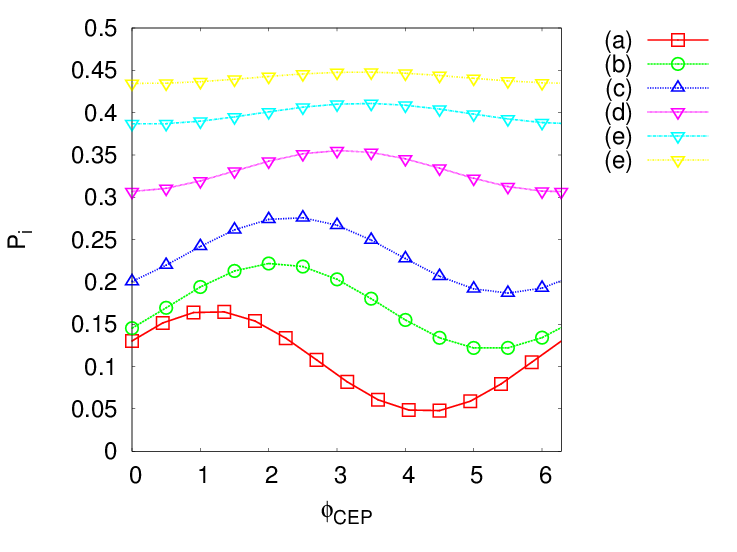}
\caption{The CEP dependence of the final
ionization probability for $\tau=500,$ $750$, $1000$, $1500$,  $2000$ and $2500$ atomic units [(a), (b), \ldots (f)]. (The corresponding intensity FWHM values are $4.4,$ $6.6,$ $8.8,$ $13.2,$ $17.6,$ $22.0$ $\mathrm{fs},$ respectively.) The initial state is $\psi_{{23}}$ with $c_{2s}=c_{3p}=1/\sqrt{2}.$ Additional parameters: $800\,\mathrm{nm}$ central wavelength, and $E_{0}=2.5\, \mathrm{GV/m}.$ }%
\label{fig5}%
\end{figure}

\subsection*{Superpositions of stationary states with different weights}
\begin{figure}[ptb]
\centering
\includegraphics[width=8cm]{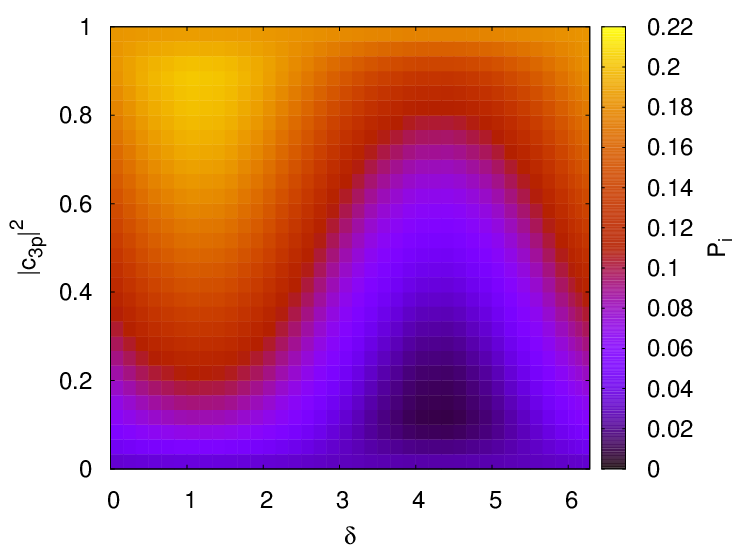}
\caption{The final ionization probability for the state $\psi_{{23}}$ as a function of the weight of the state $\phi_{{3p}}$ in the superposition and the relative phase $\delta.$}%
\label{fig6}%
\end{figure}

According to the previous subsection, the relative phase $\delta$ between the
constituents of the initial superposition, as well as the CEP of the laser pulse
can strongly modulate the ionization yield. However, the mean ionization probability,
around which oscillations appear as function of $\delta$ (or the CEP) has not been analysed so far.
As a guideline, it is reasonable to assume that the mean ionization probability mentioned above is
determined by the structure of the energy levels. On the other hand, recalling Fig.~\ref{fig3}, we see the failure
of the naive idea that the closer the expectation value of $H_0$ to the limit of the continuum (zero level of the energy) in a given
initial state, the higher the ionization yield is. (The mean ionization probability is five percent higher for
$\psi_{{34}}$ than for $\psi_{{45}}$). By inspecting the level scheme again, we see that the
states with principal quantum number $n=3$ are almost resonant with the limit of the continuum (the corresponding wavelength
is $9/R_\infty\approx 820 \,\mathrm{nm}.$) Let us recall \cite{FaisalScazano92,KopoldBecker02} that for ionization from a stationary state with energy $\epsilon_{n}$ the positions of the ATI peaks can be given by $E_m=m\hbar\omega-U_{\mathrm{p}}-|\epsilon_{n}|,$ where the ponderomotive energy $U_{\mathrm{p}}$ is practically negligible at the laser intensity we considered. That is, the first, strongest ATI peak (corresponding to $m=1$) is practically at the limit of the continuum for $n=3,$ thus the ionization probability is exceptionally high when the atomic system can be characterized
initially by the principal quantum number $n=3.$ As it is shown by Fig.~\ref{fig3}, this increased ionization probability is
visible even for a superposition when one of the states corresponds to $n=3.$ In order to see how atomic coherence phenomena
superimpose on this effect, in Fig.~\ref{fig6} we plotted the final ionization probability of the initial state $\psi_{{23}}$
as a function of both $\delta$ and the weight of the state $\phi_{{3p}},$ i.e., $|c_{3p}|^2$. If the process was completely
incoherent the final ionization probability would linearly interpolate between its values valid for the stationary states
$\phi_{{2s}}$ (corresponding to $c_{3p}=0$) and $\phi_{{3p}}$ (when $c_{3p}=1$).  The oscillations as a
function of $\delta$ are clear signatures of effects related to the atomic coherence.

\section*{Possible experimental realization}
\label{expsec}
The results of the previous sections were obtained by assuming idealized circumstances, without assessing experimental possibilities for the realization of the scheme. Let us now take this aspect into account.

The method we propose is the following: The atomic gas, which is initially in its ground state, is excited by a relatively weak, picosecond pulse (or sequence of pulses) that prepares the superpositions given by Eqs.~(\ref{psiI})-(\ref{psiIII}). Subsequently these states are probed by the ionizing few-cycle femtosecond pulse. The relative phase $\delta$ between the constituents of the superpositions (\ref{psiI})-(\ref{psiIII}) at the onset of the fs pulse can be -- in principle -- controlled by appropriately choosing the delay between the preparing pulse and the fs one. In order to obtain the $\delta$ dependence of the ionization probability (see Fig.~\ref{fig3}), several experimental runs are required, with different time delays between the pulses.
The natural time scale on which $\delta$ changes is given by the inverse of the Bohr frequencies $\omega_{nn'}$ [see Eq.~(\ref{evol0})], thus the delay between the preparing and the probe pulse should be controlled with a subfemtosecond precision. Additionally, the preparation process has to be robust enough to produce superpositions with stable phase $\delta$. The first (delay) requirement can obviously be satisfied, pump-probe experiments with attosecond delay accuracy can be performed routinely \cite{KrauszRevModPhys09}.

Let us now focus on the constancy of $\delta$ during a single experimental run.
First of all, we should consider a Doppler-free setup, to avoid inhomogeneous broadening.
 Even in this case, one might think that since different atoms unavoidably have different initial phases (i.e., initially we have $|1s\rangle\exp(i\theta),$ with $\theta$ being uniformly distributed between zero and $2\pi$), the value of $\delta$ will also have a wide distribution. However, due to the linearity of quantum mechanics, the phase term $\exp(i\theta)$ will be a multiplicative factor during the whole process, i.e., it means a global phase that plays no role in the measurement of physical observables. Similarly, phenomena that ruin quantum mechanical coherence can also be neglected, since their time scale (which is essentially the same as the lifetime of the excited levels) is much longer than the duration of a single experimental run.

The issue which is still to be investigated is if the preparation procedure itself affects the reproducibility 
of a stable phase difference $\delta$.
 Since the preparation is a long process, its full quantum mechanical treatment (using the same numerically exact methods that we applied for the ionization process) is computationally prohibitively expensive. However, due to its low intensity, only a few atomic levels play non-negligible role, thus a much simpler calculation that uses the expansion of the wave function on a few energy eigenstates is sufficient. More precisely, a multilevel model including a pair of resonant states can be reliably used to determine the experimental requirements we are interested in. Using a 500-cycle pulse with peak electric field strength having the order of magnitude of MV/m, our simple model-calculations (without the rotating wave approximation) reproduce the predictions of the area theorem \cite{AE75} almost exactly. [That is, the coefficients $c$ in Eqs.~(\ref{psiI})-(\ref{psiIII}) can be controlled precisely by changing either the peak intensity or the duration of the pulse.] Additionally, in complete accordance with textbook results, no additional phase (apart from the one that oscillates with the Bohr frequency) is gained during the preparation process -- at exact resonance. This means that a few percent change of the peak intensity or the duration of the preparing pulse induces a few percent change in the coefficients $c,$ but does not affect the relative phase $\delta.$ On the contrary, if the carrier frequency $\nu$ of the preparing pulse is not exactly the same in the consecutive pulses, $\delta$ will also have a distribution with nonzero width. According to our calculations, if one wants to keep the uncertainty of $\delta$ below 0.1 rad, the relative deviation $\Delta \nu/\nu$ should be kept below $10^{-5}.$ This experimental requirement is easily fulfilled. Note that averaging $\delta$ over the realistic interval of 0.2 rad results in a less than 5\% relative decrease in the amplitude of the oscillations in Fig.~\ref{fig3}. Additionally, effects that are related to the population of the energy levels (and not to their coherence) should be detectable even when  $\delta$ is completely random. Particularly, the mean ionization probabilities that correspond to the different curves in Fig.~3 do not change when $\delta$ becomes random.

In summary, the analysis above shows that the experimental techniques that are required to verify our theoretical findings are already available.

\section*{Conclusions}
\label{conclusionsec}
We studied the interaction of a hydrogen-like atom in different superposition
states and a few-cycle, near
infrared laser pulse. The initial atomic states were assumed to be a
superposition of \textit{s} and \textit{p} states, with different principal quantum numbers. The
ionization probability was calculated, and it was found to depend on both the
complex amplitude of the constituents of the superpositions and the waveform of
the laser pulse. This effect is the strongest when the carrier frequency of
the exciting pulse is nearly resonant with the Bohr frequency corresponding to
the initial superposition. The origin of this effect -- in a simplified, but
intuitive picture -- is that the "swing" of the internal dipole oscillations
can be excited both constructively and destructively, depending on the
relative phase of the internal, atomic oscillations and the laser field. For
more detuned excitations, this effect was shown to disappear. Furthermore, we
found that $2s-3p$ superposition states enable measurable CEP dependence in
the final ionization probability even for pulses as long as $22\,\mathrm{fs}$,
corresponding to 8 optical cycles. This is not the case for energy eigenstates,
representing a substantial difference between these two initial conditions.





\section*{Acknowledgements (not compulsory)}

We acknowledge support from the Hungarian Academy of Sciences
(“Lend\"ulet” Grant). This work was also partially supported by
the European Union and the European Social Fund through project
entitled ”ELITeam at the University of Szeged”, and by the
National R\&D Office of Hungary under contracts No. 81364 and 109257.

\section*{Author contributions statement}

V.A. performed the theoretical formulation, numerical simulation and the data collection. M.G.B. initiated the concept of the manuscript.  V.A. and F.P. conducted the numerical simulations. V.A. prepared the figures. V.A., M.G.B., P.D. and P.F. analysed the results. V.A. and P.F wrote the main manuscript text. All authors reviewed the manuscript.

\section*{Additional information}


\textbf{Competing financial interests} 

The authors declare no competing financial interests.

\end{document}